\documentclass[aps,prl,twocolumn,superscriptaddress]{revtex4-2}
\usepackage[utf8]{inputenc}
\usepackage{xcolor}
\usepackage{graphicx}
\usepackage[colorlinks,citecolor=blue,urlcolor=blue,linkcolor=blue]{hyperref}
\usepackage{mathtools}
\usepackage{amsmath}
\usepackage{amssymb}
\usepackage{braket}
\usepackage{tikz}
\usetikzlibrary{calc}

\newcommand{\eff}{\mathrm{eff}}
\renewcommand{\i}{\mathrm{i}}
\newcommand{\e}{\mathrm{e}}

\newcommand{\freiburg}{Physikalisches Institut, Albert-Ludwigs-Universität Freiburg,\\ Hermann-Herder-Straße 3, D-79104 Freiburg, Germany}
\newcommand{\eucor}{EUCOR Centre for Quantum Science and Quantum Computing, Albert-Ludwigs-Universität Freiburg,\\ Hermann-Herder-Straße 3, D-79104 Freiburg,  Germany}

\begin{document}
\title{Hong-Ou-Mandel interference of composite particles}

\author{Mama Kabir Njoya Mforifoum}
\affiliation{\freiburg}
\affiliation{\eucor}

\author{Andreas Buchleitner}
\affiliation{\freiburg}
\affiliation{\eucor}

\author{Gabriel Dufour}
\affiliation{\freiburg}
\affiliation{\eucor}

\begin{abstract}
We study the Hong-Ou-Mandel interference of two identical, composite particles, each formed of two bosonic or fermionic constituents, as they scatter against a potential barrier in a one-dimensional lattice. For tightly bound composites, we show that the combination of their constituents' mutual interactions and exchange symmetry gives rise to an effective nearest-neighbour interaction between composites, which induces a reduction of the interference contrast.   
\end{abstract}
\maketitle

The indistinguishability of identical quantum particles 
has far-reaching consequences for their dynamics. 
The most elementary---yet spectacular---example thereof is the Hong-Ou-Mandel (HOM) interference of two bosons at a balanced beam splitter \cite{hong_measurement_1987,bouchard_twophoton_2021}, where, against classical expectations, the particles always bunch into the same output.
Many-particle interference is, however, a fragile phenomenon. It is suppressed if the particles carry partially distinguishable internal states  \cite{tichy_entanglement_2011,tichy_fourphoton_2011,ra_nonmonotonic_2013,shchesnovich_sufficient_2014,tichy_sampling_2015,shchesnovich_partial_2015,rohde_boson_2015,dittel_waveparticle_2021,jones_distinguishability_2022}, or if they interact \cite{andersson_quantum_1999,longo_hongoumandel_2012,gertjerenken_effects_2015,mullin_interference_2015,dufour_manyparticle_2017,yannouleas_interference_2019,njoya}.
Here, we want to investigate another departure from the ideal HOM scenario by considering the interference of composite---rather than elementary---particles.

Intuitively, one expects that quantum effects such as many-particle interference are suppressed as the size of the compounds grows (see \cite{eibenberger_matter_2013} for the \textit{single-particle} interference of compounds).
Our aim will be to explore the precise mechanisms behind this quantum-to-classical transition. 
In other words, we ask what it takes for a composite object to display---or not---bosonic or fermionic features.
This question is often reduced to a simple counting problem, with the number of fermionic constituents fixing the nature of the composite, as suggested by the spin-statistics theorem \cite{fierz_uber_1939,pauli_connection_1940,pauli_connection_1950}.
Such a reasoning not only fails to consider the internal degrees of freedom of the composites,  which might render them distinguishable,
but it also takes the existence of the compounds for granted.
Going one step further, one can argue that composites will behave as elementary objects on energy scales well below their binding energy, such that interactions between constituents must play a central role. At this point, one should pause to consider that interactions between the constituents also imply interactions between the composites, whereas these are absent from the original HOM scenario. Furthermore, as pointed out in \cite{combescot_effective_2002,combescot_manybody_2005,combescot_shiva_2007,combescot_manybody_2008,combescot_role_2009}, the behaviour of composite particles is also affected by processes where indistinguishable constituents are exchanged between them.

It is therefore clear that many physical ingredients may play a role in the HOM interference of two composites. To complete the picture, it has been suggested that entanglement between the constituents, rather than interaction, is essential for the emergence of bosonic statistics in two-fermion composites \cite{law_quantum_2005,chudzicki_entanglement_2010,ramanathan_criteria_2011,combescot_commutator_2011,tichy_collective_2012,gavrilik_entanglement_2012}. 
These formal investigations have initially focused on deviations from the ideal bosonic ladder describing the occupation of a single mode (see also \cite{combescot_new_2001,combescot_nexciton_2003}), but later works have also considered interference scenarios, which were, however, specifically designed to accommodate the desired entanglement structure
\cite{kurzynski_particle_2012,tichy_collective_2012,lasmar_nonlocal_2017}.

In this contribution, we adopt a more direct approach, setting the interference scenario first, and only then identifying the physical mechanisms which explain the outcome. We consider a one-dimensional (1D) tight-binding lattice on which elementary particles are bound into composites by a contact interaction. 
As a first step away from the quantum limit, we consider composite bosons (cobosons) made up of two elementary constituents, either bosons or fermions. We then study the HOM interference of two identical cobosons which scatter on a potential barrier playing the role of the beam splitter, building on our work on the HOM interference of elementary particles on a lattice \cite{njoya}. 
We focus on the bunching probability---i.e. the probability that both composites end up on the same side of the barrier---as a figure of merit for bosonic behaviour.
Indeed, for elementary particles, the bunching probability of bosons is twice the classical value (obtained for distinguishable particles), while it identically vanishes for fermions.

The composite particles which we consider are made up of two constituents $A$ and $B$ which are distinguishable from one another.
Letting $\ket{l,l'}=\ket{l}_A\otimes\ket{l'}_B$ denote the state where particle $A$ is located on site $l$ and particle $B$ on site $l'$,
 the two-particle Hamiltonian reads
\begin{equation}\label{eq:H2}
 H_2 = H_1\otimes\mathbb{I}_1 + \mathbb{I}_1\otimes H_1 + U \sum_l \ket{l,l}\bra{l,l}~,
\end{equation}
where the single-particle Hamiltonian 
\begin{equation}\label{eq:H1}
	H_1=-J \sum_{l} \Big( \ket{l}\bra{l+1}+\ket{l+1}\bra{l} \Big) +\mu \ket{0}\bra{0}
\end{equation}
describes both tunnelling between neighbouring sites (with a coupling constant $J>0$) and a potential barrier of height $\mu$ on site $l=0$. In Eq.~\eqref{eq:H2}, $\mathbb{I}_1=\sum_l \ket{l}\bra{l}$ is the identity on the single-particle Hilbert space and $U$ is the strength of the contact interaction which will allow to bind the particles. 

The scattering and bound eigenstates of $H_2$ are described in \cite{njoya}. 
For sufficiently large values of $|U|/J$, the latter form a band with energies $E_\mathrm{b}\approx U$, 
suggesting  that an effective description of the bound pair as a single object is possible.
To derive an effective Hamiltonian for such a tightly bound pair, we focus on the subspace spanned by paired states $\ket{l,l}$ and consider their coupling through second-order processes via an unpaired intermediate state $\ket{l,l'}, \, l\neq l'$.
For example, the state $\ket{l,l}$ is coupled to $\ket{l+1,l+1}$ by two tunnelling events, via either one of two possible intermediate states: $\ket{l+1,l}$ and $\ket{l,l+1}$.
The corresponding matrix element of the effective Hamiltonian is given by the product of the two tunnelling amplitudes, divided by the interaction-energy difference between paired and unpaired states, summed over all possible intermediate states \cite{cohen-tannoudji_atomphoton_1998}.
This results in an effective tunnelling constant $J_\eff = -2J^2/U$ for the composite. 
In addition, the state $\ket{l,l}$  is also coupled to itself via four possible intermediate states  (each of the two particles can tunnel to the left or right and back), resulting in an effective energy shift $4J^2/U=-2 J_\eff$. 
Since this effect is uniform in the bulk of the lattice, it only contributes a constant energy shift on top of the interaction energy $U$ and plays no role at this stage. Up to these constants, the effective Hamiltonian obtained with this approach thus has the same form as the single-particle Hamiltonian \eqref{eq:H1}, with the replacements $\ket{l}\to \ket{l,l}$, $J\to J_\eff$ and a doubling of the barrier height $\mu\to 2\mu$ \cite{compagno_noon_2017}. Note that for $U>0$, the composite is repulsively bound  \cite{winkler2006repulsively} and $J_\eff <0$, so that the dispersion relation of the composite is sign-flipped  compared to that of a single particle.

\begin{figure}

	\begin{tikzpicture}
		\draw (0,0) node{\includegraphics[width=7cm]{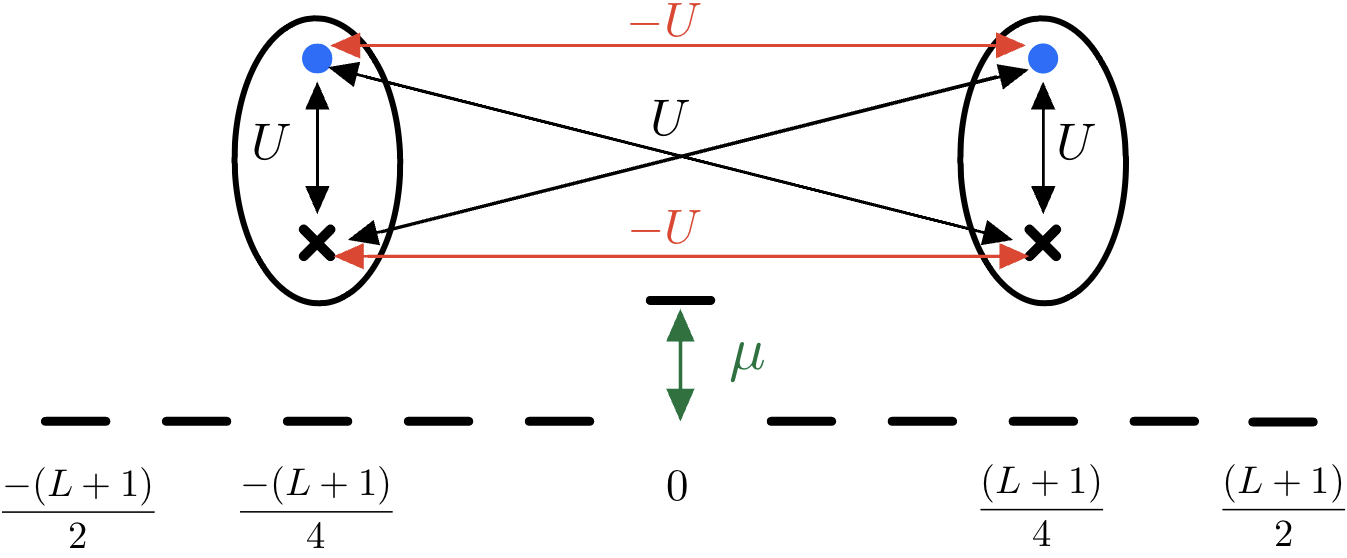}};
		\draw (-3.8,-1.23) node{\scriptsize$l=$};
	\end{tikzpicture}
	
	\caption{Two composite particles on a $L$-site 1D lattice with a potential barrier of height $\mu$. The black crosses and blue dots represent the constituents of type $A$ and $B$, respectively. The on-site interaction strength between constituents of the same type is opposite to that between constituents of different types.}
	\label{fig:lattice}
\end{figure}

We now consider two particles of type $A$ and two of type $B$, assembled into two $AB$ composites, as sketched in Fig.~\ref{fig:lattice}. 
States where each particle is located on a well-defined lattice site are denoted
$\ket{l,l',m,m'}=\ket{l}_A\otimes\ket{l'}_B\otimes\ket{m}_A\otimes\ket{m'}_B$.
The identity of the two composites imposes that the interaction be the same between any two particles of different types, i.e. the contact interaction $U$ used to form the composites.
However, we are free to choose the interaction between particles of the same type, and we choose a contact  
 interaction of strength $-U$ (see Fig.~\ref{fig:lattice}), such that the total interaction energy remains equal to $2U$ whether the two composites are located  on the same site or on different sites.
 With the above definition of the four-particle Hamiltonian, we hope to reproduce as faithfully as possible the behaviour of non-interacting particles. 

In the limit $|U|\gg J$, we can again derive an effective Hamiltonian, now for two composite particles. We restrict the four-particle Hilbert space to the subspace where the particles are assembled into two composites, each consisting of an $A$ and a $B$ particle, i.e., the subspace with interaction energy $2U$. It is spanned by two sets of states, 
$\{\ket{l,l,m,m}\}_{l,m}$ and $\{\ket{l,m,m,l}\}_{l,m}$, depending on which $A$ particle is paired with which $B$ particle. States  $\ket{l,l,l,l}$, where all four particles are on the same site, form the intersection of these two sets. The configuration space of two tightly bound composites therefore has a more complex structure \cite{njoyaPhD} than that of two elementary particles, which consists of a single 2D square lattice $\{\ket{l,m}\}_{l,m}$, as described in \cite{njoya}. 
The derivation of the effective Hamiltonian's matrix elements essentially follows that of the single-composite case:
the effective coupling for moving either one of the two composites to the left or to the right is given by $J_\eff=-2J^2/U$ and all states receive a global energy shift, which is twice that of the single-composite case, i.e. $2(U-2 J_\eff)$.
However, when the composites are located on neighbouring sites, the non-trivial structure of the configuration space comes into play: the composites can additionally exchange their $A$ or their $B$ constituents, also with an effective coupling $J_\eff$. These exchange processes connect the two families of states, $\{\ket{l,l,m,m}\}_{l,m}$ and $\{\ket{l,m,m,l}\}_{l,m}$.
For now, we have not taken the bosonic or fermionic nature of the constituents into account. As we will show in the following, enforcing these symmetries erases the distinction between the two aforementioned families of states but instead gives rise to exchange interactions between the composites.

To see this, we introduce the operator $E_A: \ket{l,l',m,m'}\mapsto\ket{m,l',l,m'}$ which exchanges the particles of type $A$, the corresponding operator $E_B:\ket{l,l',m,m'}\mapsto\ket{l,m',m,l'}$, as well as the parity operator $P:\ket{l,l',m,m'}\mapsto\ket{-l,-l',-m,-m'}$ which reflects the lattice around its centre.
These are seen to commute among themselves and with both the full four-particle Hamiltonian and the corresponding two-composite effective Hamiltonian.
The Hilbert space thus decomposes into eight decoupled symmetry sectors, each associated with a triplet $(\epsilon_A,\epsilon_B,\pi)$ of eigenvalues $\pm1$ of the symmetry operators.
If the particles of type $A$ are indistinguishable bosons, only sectors with $\epsilon_A=1$ can be populated, whereas  $\epsilon_A$ must be $-1$ for indistinguishable fermions. Likewise, the nature of the $B$ particles fixes $\epsilon_B$.
 For fixed values of $(\epsilon_A,\epsilon_B,\pi)$, a state of two tightly bound composites is fully defined by the coefficients of the symmetry-adapted basis states $(\mathbb{I}+\epsilon_A E_A)(\mathbb{I}+\epsilon_B E_B)(\mathbb{I}+\pi P)\ket{l,l,m,m}$ for $l\leq 0$ and $-|l|\leq m \leq |l|$. This triangular portion of the  $lm$-plane defines the reduced configuration space of two identical tightly bound composites and is the same as that of two elementary indistinguishable particles described in \cite{njoya}.
   Following the approach developed there, we can express the bunching probability of two composites symmetrically launched towards the central barrier as
\begin{equation}\label{Pb}
	P_\mathrm{b}=|RT+T\e^{\i\phi}R|^2~,
\end{equation}
where $T$ and $R$ are the transmission and reflection coefficients of the barrier while $\phi$ is the scattering phase shift accumulated if the particles collide. 
This expression can be interpreted as describing the interference of two indistinguishable sequences of events, each involving the reflection of one particle and the transmission of the other one:
 either the reflection occurs \textit{before} the transmission, such that the particles \textit{do not} meet, or the reflection takes places \textit{after} the transmission, such that the particles \textit{do} meet and a phase shift $\phi$ is applied.
For non-interacting elementary bosons, $\phi=0$, yielding the bunching probability 	$P_\mathrm{b}^\mathrm{elem}=4|TR|^2$, which reaches unity for a balanced beam splitter $|T|^2=|R|^2=1/2$.
A finite phase shift $\phi$, be it due to interaction between elementary particles, as in \cite{njoya}, or, as we will discuss now, to the composite nature of the interfering objects, always entails a reduction of the bunching probability.


The interaction between two identical composite particles and the accompanying scattering phase shift $\phi$  arise because of exchange contributions to the effective couplings and energy shifts.
 Consider for example composites located on neighbouring sites. For fermionic constituents, because of Pauli's exclusion principle, the constituents of one composite cannot tunnel to the site occupied by the other composite.
For bosonic constituents, on the other hand, this is not only possible, but there are two ways of returning to the original configuration: either the same particle hops back, or its indistinguishable partner does. As a result, the configurations where the composites occupy neighbouring sites receive an effective energy shift which depends on the quantum statistics of the constituents.

\begin{figure}
	\centering
	\includegraphics[width=0.35\textwidth]{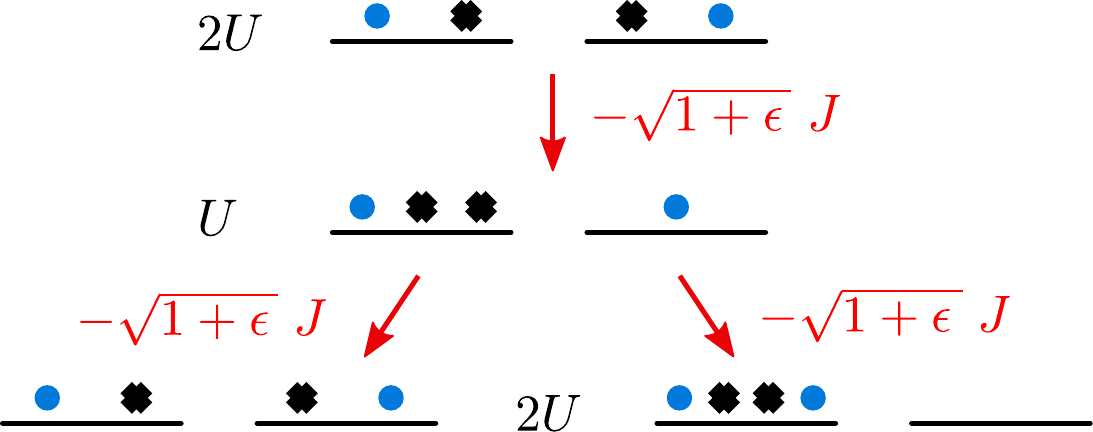}
	\caption{Second order processes contributing to the effective Hamiltonian of two tightly bound composite particles. A state where the composites occupy neighbouring sites (top)---with interaction energy $2U$---is coupled to itself (bottom left)  and to the state where all particles occupy the same site (bottom right)---both with the same total interaction energy $2U$---via an intermediate state of energy $U$ (centre). The corresponding transition amplitudes, shown in red, are enhanced for bosonic constituents ($\epsilon=1$) and suppressed for fermionic ones ($\epsilon=-1$).}
	\label{fig:effective}
\end{figure}

To account for such exchange contributions, we evaluate the effective Hamiltonian in the symmetry-adapted basis. In the following, we take $\epsilon_A=\epsilon_B=\epsilon$ and consider cobosons made up of either two fermions ($\epsilon=-1$) or two bosons ($\epsilon=1$).
In Fig.~\ref{fig:effective}, we sketch second-order processes contributing to 
	(i) the effective energy shift of configurations where composites occupy neighbouring sites and (ii) starting from such a configuration, the effective tunnel coupling to a state $\ket{l,l,l,l}$ where both composites occupy the same site. 
For fermionic constituents   ($\epsilon=-1$),
	transition amplitudes to states where two particles of the same type occupy the same site  vanish.
	For bosonic constituents ($\epsilon=1$), on the contrary, such transition amplitudes are enhanced by a factor $\sqrt{2}$.
	The tunnelling amplitude to or from a state where two identical particles occupy the same site thus reads $-\sqrt{1+\epsilon}J$, as indicated in Fig.~\ref{fig:effective}.
The consequences of this modified single-particle tunnelling amplitude are 
(i)  an effective energy shift $ -(4+2\epsilon) J_\eff$ for composites located on neighbouring sites, differing by $-2\epsilon J_\eff$ from the effective energy shift of composites located further apart from each other,
(ii) an effective tunnel coupling $-(1+\epsilon)J_\eff $ for transitions to and from  states $\ket{l,l,l,l}$.
In addition to the fermionic suppression or bosonic enhancement of transitions to states where all particles occupy the same site, identical composites thus experience a nearest-neighbour interaction of strength $-2\epsilon J_\eff$ arising from the exchange symmetry of their constituents.
One can draw a parallel to the exchange interaction \cite{heisenberg_mehrkoerperproblem_1926,dirac_theory_1926,heisenberg_zur_1928,dirac_quantum_1929,vanvlek_theory_1932,anderson_theory_1963} between electrons 
occupying neighbouring sites in the Mott-insulating phase of the Hubbard model
provided their spins are aligned (making them indistinguishable) and 
giving rise to a coupling between the orientations of neighbouring spins \cite{cleveland_obtaining_1976,anderson_new_1959}. Here, however, the particles are not pinned to a lattice site, but are bound in pairs which can propagate and scatter off one another.

With the boundary conditions described above, we can calculate the scattering phase $\phi$ between two composites with a relative quasimomemtum $k$ by solving the corresponding Schrödinger equation.
We find $\e^{\i\phi}=-\e^{\i k}$
for cobosons made of fermions ($\epsilon=-1$) and 
$\e^{\i\phi}=\e^{\i k}(\sin k -\i)/(\sin k +\i)$ for  cobosons made of bosons ($\epsilon=1$).
The bunching probability for composites is thus given by  $P_\mathrm{b}^\mathrm{comp}=s(k) P_\mathrm{b}^\mathrm{elem}$, with a suppression factor due to compositeness
\begin{equation}
	s(k)=\begin{cases}
		\sin^2\dfrac{k}{2} & \text{for fermions,}\\
		\dfrac{(2+\cos k)^2 }{1+\sin^2 k}\sin^2\dfrac{k}{2} & \text{for bosons.}
	\end{cases}\label{eq:suppr}
\end{equation}
In both cases, $s(k)\leq 1$ depends only on the relative quasimomentum $k$, and it vanishes for $k\to0$ and goes to one for $k\to\pi$. 
While $s(k)$ monotonously increases between these two limits in the fermionic case, this is not true of the bosonic factor, which also reaches one for $k=\pi/2$. 
The above calculation assumes infinitely extended wavepackets, i.e. plane waves, 
 and neglects the formation of bound states of the composites with the barrier or with each other (see the discussion in \cite{njoya}). Despite these limitations, it still gives a good estimate of the bunching probability, as we now demonstrate numerically.

We simulate the dynamics of the composites by diagonalizing the effective two-composite Hamiltonian described above. 
 Comparison with the dynamics generated by the full four-particle Hamiltonian shows qualitative agreement for interaction energies $|U|$ above $5J$, and quantitative agreement above $20J$. The effective approach allows to reduce the dimension of the problem from $L^4$ to  $2L^2-L$, and we can further divide this number by eight using the exchange and parity symmetries. We can therefore push the lattice size to $L=101$ in the following simulations.
The composite particles are prepared in Gaussian wave packets placed symmetrically on both sides of the barrier, with opposite quasimomenta. The initial state thus reads
\begin{equation}
\begin{split}
	&\ket{\psi_0}=\sum_{l,m} G_{c,k,\sigma}(l) G_{-c,-k,\sigma}(m) \ket{l,l,m,m}~,\\
	&\text {with} \quad G_{c,k,\sigma}(l)= \exp\left[ (l-c)^2/4\sigma^2+ikl\right]~. \label{eq:init}
\end{split}
\end{equation}
We choose the width $\sigma=10$ of the wavepackets small with respect to their distance $|c|=(L+1)/4$ from the barrier, such that the wave packets are initially non-overlapping,
but large enough with respect to $1$, so that the quasimomentum is sharply defined.
We restrict the values  of the quasimomentum $k$ to the range $[\pi/6,5\pi/6]$, such that the wave packets propagate with a finite group velocity $v_g= 2\hbar^{-1}  J_\eff \sin k$ and limited dispersion.
We additionally apply the  symmetrization operator
$(\mathbb{I}+\epsilon E_A)(\mathbb{I}+\epsilon E_B)$ to Eq.~\eqref{eq:init},
according to the nature of the constituents, before normalizing.
The bunching probability is evaluated after an evolution time $t=(3L+1)/4|v_g|$, which allows the wave packets to reach the barrier and then propagate away towards the boundaries of the lattice.

\begin{figure}
	\centering
    \includegraphics[width=0.49\textwidth]{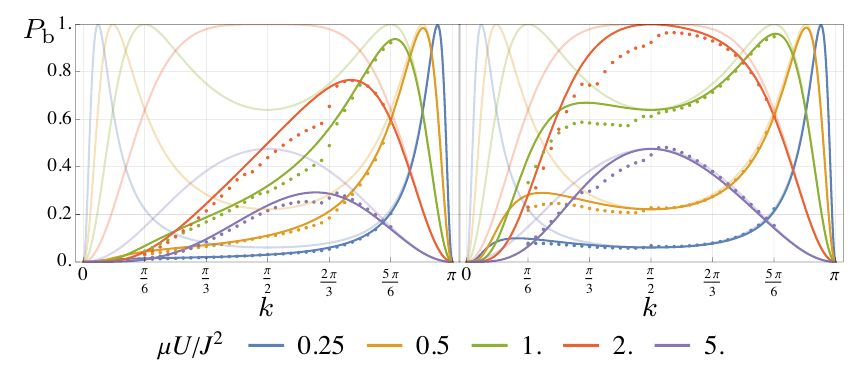}

	\caption{
		Bunching probability of composite particles
		made of fermions (left) or bosons (right) as a function of the quasimomentum $k$, for various values of $\mu U/J^2$ (see legend).
		The full lines show the analytical predictions $P^\text{comp}_\mathrm{b}$ for composite bosons. 
		For comparison, the corresponding value $P^\text{elem}_\mathrm{b}$  for elementary particles are shown as faint lines.
		The points show the results of the numerical wave-packet propagation with the effective Hamiltonian.}
	
	\label{fig:bunching}
	
\end{figure}

In Fig.~\ref{fig:bunching}, we show the bunching probability of cobosons 
made of fermions (left panel) or bosons (right panel) in our simulations (dots), alongside  the theoretical predictions for elementary particles 
\begin{equation}
	P_\mathrm{b}^\mathrm{elem}=\frac{16 (\mu U/J^2) ^2 \sin^2 k}{ \left(4 \sin^2 k+(\mu U/J^2) ^2\right)^2}
\end{equation}
(faint lines, 
obtained by making the replacements $J\to J_\eff$ and  $\mu \to 2\mu$ in the corresponding expression of
\cite{njoya}) and for composites $P_\mathrm{b}^\mathrm{comp}=s(k) P_\mathrm{b}^\mathrm{elem}$ (bright lines).
The latter are seen to agree well with simulation results over the considered parameter range, with deviations which we attribute to composites binding to the barrier and/or to each other.

These results confirm that the exchange of constituents between composite particles---which gives rise to an effective interaction and an associated phase-shift---constitutes the main impact of compositeness on many-particle interference in the present scenario.
Interestingly, for both bosonic and fermionic constituents,
the phase shift goes to $\pi$ in the limit of small quasi-momenta, resulting in a vanishing bunching probability for the cobosons, whereas it goes to zero in the limit $k\to\pi$ (and, for bosons, also at $k=\pi/2$), such that the bunching probability of composites approaches that of elementary particles. This is in stark contrast with the $k\to\pi-k$ symmetry observed in the behaviour of elementary bosons.
These limits are difficult to approach in practice given that the group velocity vanishes. Nevertheless, we were able to numerically observe bunching probabilities of up to 92\% for cobosons made of fermions (for $\mu U=J^2$ and $k=5\pi/6$).

While the quality of bosonic interference depends very much on the quasimomentum $k$ (which controls the scattering of the cobosons), entanglement between the constituents seems to play at most a marginal role in the present setup, in contrast to other scenarios \cite{law_quantum_2005,chudzicki_entanglement_2010,ramanathan_criteria_2011,combescot_commutator_2011,tichy_collective_2012,gavrilik_entanglement_2012,kurzynski_particle_2012,tichy_collective_2012,lasmar_nonlocal_2017}.
In the initial state \eqref{eq:init}, the entanglement between the positions of the two constituents of a composite is set by the extension $\sigma$ of the center-of-mass wave packet \cite{lasmar_dynamical_2018}. It is therefore related to how sharply the center-of-mass momentum is defined around its central value $k$, but is independent of $k$ itself. Moreover, given that the particles interact, this entanglement is not conserved during the evolution. 

Finally, it is noteworthy that the observed effect of compositeness persists in the limit of infinite binding energy $E_\text{b}\approx U$ of the compounds, barring a naive approach to the problem in  terms of a separation of energy scales. Indeed, the exchange interactions behind the deviations from elementary behaviour remain of the same order as the effective tunnel coupling $J_\eff$ of the composites and are therefore always relevant for their dynamics.

\begin{acknowledgements}
\textit{Acknowledgements---}	
The authors are indebted to Shahid Iqbal for stimulating discussions on the connection between composite behavior and entanglement. 
We acknowledge support by the state of Baden-Württemberg through bwHPC
and the German Research Foundation (DFG) through grant no INST 40/575-1 FUGG (JUSTUS 2 cluster). 
MKNM acknowledges support by the German Research Foundation (DFG) through research training groups IRTG 2079 CoCo and RTG 2717 Dyncam.
\end{acknowledgements}

\bibliography{HOMcomposites}

\end{document}